# Role of Fermi-level pinning in nanotube Schottky diodes

François Léonard and J. Tersoff
*IBM Research Division, T. J. Watson Research Center, P.O. Box 218,
Yorktown Heights, NY 10598*
(March 21, 2000)

At semiconductor-metal junctions, the Schottky barrier height is generally fixed by "Fermi-level pinning". We find that when a semiconducting carbon nanotube is end-contacted to a metal (the optimal geometry for nanodevices), the behavior is radically different. Even when the Fermi level is fully "pinned" at the interface, the turn-on voltage is that expected for an *unpinned* junction. Thus the threshold may be adjusted for optimal device performance, which is not possible in planar contacts. Similar behavior is expected at heterojunctions between nanotubes and semiconductors.

73.61.Wp, 85.30.Vw, 73.30.+y, 73.40.Ns

Metal-semiconductor junctions play a crucial role in electronic devices. They are useful as active device elements (Schottky diodes), or simply to provide electrical contact to a semiconductor device. The earliest model suggested that any desired barrier height (including ohmic contacts) could be obtained by using a metal of appropriate workfunction [1]. Unfortunately, the actual junction is almost invariably a Schottky barrier, with the Fermi level at the interface "pinned" deep in the semiconductor bandgap, regardless of the metal workfunction. This places a serious constraint on device design.

For nanoscale devices based on carbon nanotubes (NTs) or other linear molecules, it seems clear that contacts will play an even more central role in device performance, since the entire device may lie within nanometers of the interface. Yet little is known about what role Fermi-level pinning will play in such devices.

Most metal-NT junctions studied to date employ a weak van der Waals side contact, as in Fig. 1a [2], and the Fermi-level alignment at such contacts has recently been considered [3]. However, as discussed by Zhang *et al.* [4], in real applications an end-bonded junction with strong (metallic/covalent) bonding will be preferable for compactness and for robust electrical contact; and such junctions have already been fabricated [4]. In such junctions, one expects Fermi-level pinning just as in traditional devices.

Here we show that this Fermi-level pinning, no matter how strong, cannot control the device properties of metal-NT contacts in this geometry. In a planar geometry, pinning determines the turn-on voltage; but for the quasi-one-dimensional geometry of NTs, pinning at most creates a leaky tunnel barrier in series with the junction. The low-temperature voltage threshold remains exactly that of the *unpinned* interface. Thus the choice of metal workfunction may provide a powerful tool to control the contact behavior — a tool that is not available for planar junctions.

The theory of Schottky barriers has at times been highly controversial, with a disconcerting variety of models [5]. However, the key physics of Fermi-level pinning at planar interfaces is actually rather straightforward [6–8]. As illustrated in Fig. 2, the barrier height is $\phi_b = E_C - E_F$, where $E_C$ and $E_F$ are the energy of the conduction-band edge and the Fermi level in the barrier region near the interface. For simplicity, we restrict discussion throughout to the barrier for electrons; a trivial modification gives the barrier for holes. Also, we make the usual assumption that the depletion length $L_d$ is long compared to other length scales here, so there is negligible band-bending due to doping in the near-interface region.

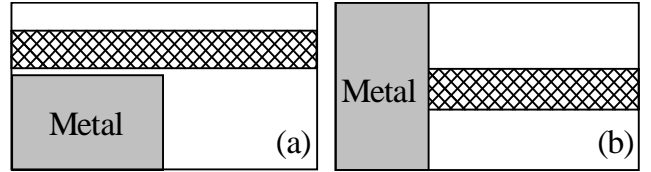

FIG. 1. Two types of nanotube/metal contacts: (a) nanotube (cross-hatched) side-contacted by the metal by van der Waals adhesion; (b) nanotube end-bonded to the metal (covalent/metallic bonding).

For a metal and semiconductor separated by a thin vacuum gap ($\ll L_d$), the barrier is simply [1]

$$\phi_{b0} = \chi_m - \chi_s , \qquad (1)$$

where $\chi_m$ and $\chi_s$ are the metal workfunction and semiconductor ionization potential. This is typically taken to be the barrier expected in the absence of an interface dipole [1]. ($\chi_m$ and $\chi_s$ depend on details of the free surface that cannot be relevant to the real planar interface, and are often replaced with values for some idealized surface [8].)

However, at a metal-semiconductor interface there is in general a finite density of states throughout the bandgap of the semiconductor near the interface, the so-called "metal-induced gap states" (MIGS) [6,7]. Like surface states, these are simply Bloch states of the semiconductor having complex wavevector. The boundary condition at the interface with a metal allows states at all energies, with states in the bandgap decaying exponentially into the semiconductor. These MIGS (together with their image charge in the metal) give a dipole $D$ at the interface. At a planar interface, this raises or lowers the barrier to

$$\phi_b = \phi_{b0} + D \ . \qquad (2)$$

The occupancy of the gap states, and hence the dipole, depends on the position of the Fermi level within the semiconductor bandgap (Fig. 2). Within a linear theory this can be written quite generally in terms of two parameters $E_N$ and $\alpha$ as

$$D = \alpha \left( E_F - E_N \right) = \alpha \left( \phi_{bp} - \phi_b \right) \ . \qquad (3)$$

Here $\phi_{bp} = E_C - E_N$ is the barrier height in the limit of strong pinning (large $\alpha$), and $E_N$ is the "neutrality level", i.e. the Fermi-level position at which the dipole would be zero. With Eq. (2), this gives a self-consistent barrier height of

$$\phi_b = \frac{1}{\alpha+1}\phi_{b0} + \frac{\alpha}{\alpha+1}\phi_{bp} \ . \qquad (4)$$

Simple arguments indicate that $\alpha + 1$ is *at least* equal to the semiconductor dielectric constant, typically $\gtrsim 10$; and this is supported by comparison with experiment [9]. Therefore $\phi_b \sim \phi_{bp} = E_C - E_N$. The position of $E_N$ in the semiconductor bandgap must in principle depend on the atomic-level structure of the interface. However, simple theories that identify $E_N$ with some generic midgap position have had considerable success [7,9,10].

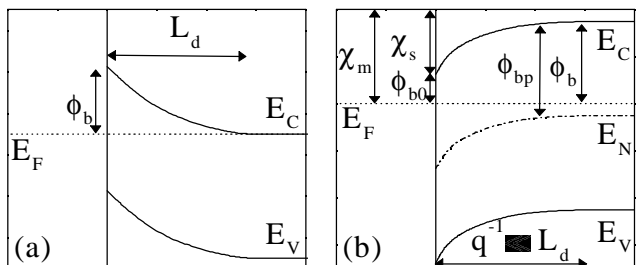

FIG. 2. Schematic band diagram for planar Schottky barrier, showing energy of Fermi level and local band edges vs. distance from interface (vertical line). (a) Large scale, showing band bending over length $L_d$ due to doping. (b) Close-up of interface region, showing dipole (spread over length $q^{-1}$) due to MIGS. See text for notation.

For NTs, we may immediately anticipate a fundamental change because the reduced dimensionality of the interface changes the screening. To calculate the behavior of NT contacts, we need a more microscopic picture of the interface dipole. However, we want a generic description, independent of the atomic details of a specific interface. We therefore consider a NT terminating at a planar metal contact, modeling the dipole by a charge

$$\sigma(z) = D_0 \left( E_N - E_F \right) e^{-qz} \qquad (5)$$

in the NT plus an image charge in the metal. Here $z$ is distance from the NT-metal interface, and $E_N$ varies with $z$ due to the electrostatic potential (as illustrated in Fig. 2b).

Equation (5) is of course a simplification, but it is consistent with the well-established picture of MIGS [6,7], and it captures the key aspects of the real interface dipole. All states in the bandgap decay exponentially with distance, but each state has its own decay length; so the overall decay is not truly exponential, and varies somewhat with energy in the bandgap. Here $q$ represents an appropriate average value. We note that a lower bound for $q$ is twice the NT $\pi$-band wavevector deep in the gap, so $q > 1$ nm$^{-1}$. From the calculations of Chico et al. [11] we can extract an effective decay constant of $q \sim 2$ nm$^{-1}$. (This may be an underestimate, since we lack information on the contribution of the most rapidly-decaying gap states.) We do not have an independent estimate of $D_0$, and in general this (like $E_N$) could be affected to some extent by atomic details of the interface (including incorporation of impurities). We therefore consider a range of values, focusing on the regime where $D_0$ is large enough to fully pin the bands. We emphasize that larger values of $q$ or smaller values of $D_0$ would only strengthen our conclusions.

We self-consistently calculate the charge (5) and resulting potential. Details are as in [12]. We use parameters for typical semiconducting nanotubes [13]: radius 0.7 nm and bandgap 0.6 eV. All calculations are for an infinite NT with light $n$-type doping (atomic fraction $10^{-4}$), but the key conclusions also hold for an undoped NT of any length $\gg q^{-1}$ with an ohmic back contact. For concreteness we take the neutrality level $E_N$ at the NT midgap, but a different choice would not affect our conclusions. In a real device, the NT would probably be embedded in a dielectric to enhance mechanical and chemical stability. We therefore include a dielectric constant of 3.9 as for SiO$_2$; but within our model this merely rescales $D_0$ and so does not affect our conclusions.

The calculated behavior at NT-metal contacts is shown in Fig. 3 for different values of the pinning strength $D_0$, with $q = 2$ nm$^{-1}$. The corresponding planar junctions are also shown for comparison. For the planar junction, the dipole is a sheet, so it shifts the semiconductor bands relative to the metal Fermi level even at "infinite" distances (i.e. distances comparable to the lateral dimension). In contrast, for the NT the dipole is localized in all three directions, so its effect on the potential decays as $z^{-2}$ at distances $\gtrsim 2$ nm.

We first consider the case of a high-workfunction metal ($\chi_m > \chi_s + \phi_{bp}$), Fig. 3a, where Fermi-level pinning tends to decrease the barrier for electrons. For a planar junction (Fig. 3a inset) the barrier height decreases with $D_0$, asymptotically approaching the value $\phi_{bp}$ (0.3 eV here) with the Fermi level at the neutrality level $E_N$.

In contrast, at the NT junction this barrier-lowering by the interface dipole is confined to a region of $\sim 2$ nm. The magnitude of the effect increases with $D_0$, shifting the bands toward the pinned position. However, even for very strong pinning, the effect is only local. At distances $\gg 2$ nm (but $\ll L_d$) from the junction there is still a barrier $\phi_{b0} = \chi_m - \chi_s$ (0.45 eV in our example) which the electron must surmount.

Thus for the NT, Fermi-level pinning has *no effect* on the Schottky barrier height in this case. Rather, the barrier height is controlled by the metal workfunction, just as if there were no pinning and no interface dipole. This opens the possibility of controlling the barrier height by the choice of metal. Even for a given metal, an atomically thin surface coating can substantially change the metal workfunction and so shift the voltage threshold of the device.

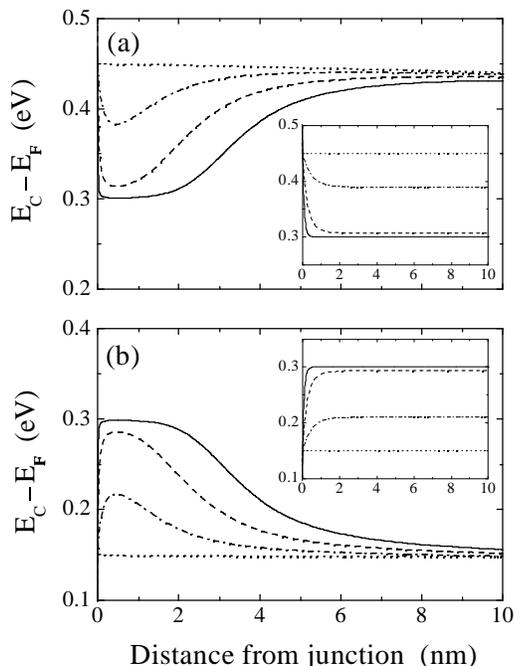

FIG. 3. Local conduction-band edge for NT as in Fig. 1b, versus distance $z$ from interface. Here $V = 0$, $q = 2$ nm$^{-1}$, nominal pinned barrier $\phi_{bp} = 0.3$ eV. (a) High-workfunction metal: $\phi_{b0} = 0.45$ eV. (b) Low-workfunction metal: $\phi_{b0} = 0.15$ eV. Dotted, dash-dotted, dashed, and solid lines are for $D_0 = 0$, 0.01, 0.1, and 1 state/(atom-eV), respectively. Insets show planar junction for same parameters (and atomic density $5 \times 10^{22}$ cm$^{-3}$).

For a low-workfunction metal we have a somewhat different situation, shown in Fig. 3b. The interface dipole raises the bands, creating an extra barrier for electrons; but again, for NTs the effect is confined to a region $\sim$ 2 nm. Electrons can tunnel through this barrier, so it decreases the current but does not affect the low-temperature threshold voltage.

For device operation, what matters most is the current as a function of voltage. We calculate this for an infinitely long tube (treated as in [12]), using approximations that are accurate in the low-current regime. A forward or reverse bias $V$ corresponds to raising or lowering the Fermi level of the NT with respect to that of the metal by an amount $eV$. The occupation of MIGS in Eq. (5) is determined by the metal Fermi level. (There is a crossover region in the NT where the local Fermi level is undefined; but the charge density in this region is negligible, so the precise treatment of the crossover does not affect the results.)

The current is given by the Landauer-Büttiker formula for a one-dimensional system [14]:

$$I(V) = \frac{4e}{h} \int P(E,V) \left[ F(E-eV) - F(E) \right] dE. \quad (6)$$

Here $P(E,V)$ is the transmission probability across the junction at bias $V$ for electrons of energy $E$, and $F(E)$ is the Fermi function for the metal. Only the lowest conduction band of the NT contributes here, giving 2 channels. The transmission probability across the barrier is calculated using the WKB approximation.

Within these approximations, a NT contact to a high-workfunction metal (Fig. 3a) is *completely unaffected* by Fermi-level pinning. Regardless of the value of $D_0$, the entire I-V curve is that expected from standard models [1] based on the *unpinned* barrier height.

The case of a low-workfunction metal is somewhat more complex. For clarity we begin with the behavior at low temperature, i.e. the low-temperature limit of Eq. (6). (We do not consider correlated-electron effects that arise very near 0K.) The tunneling probability $P$ through the local interface barrier is shown in Fig. 4a versus voltage $V$. For a doped NT, the current is well approximated by $I(V) = I_0 P(V)$, where the constant $I_0$ depends on the doping and reflects the carrier density. (For an undoped tube with ohmic back-contact, the carrier density increases with voltage so $I_0$ is no longer a constant.) We include results for three different values of $q$, to illustrate the dependence on this parameter. Since the interesting regime is that of strong pinning, for each value of $q$ we choose the pinning strength $D_0$ large enough that the top of the barrier is within 0.01 eV of its asymptotic ($D_0 \to \infty$) position.

For biases less than the *unpinned* barrier, there is no current. (The depletion region is too wide to tunnel through except at extremely high doping.) As the bias increases above the unpinned barrier height (0.15 eV in our example), electrons can tunnel from the NT to the metal though the localized barrier. Hence the device turns on at the *unpinned* barrier height, instead of the pinned barrier height as in a planar junction. Slightly above this threshold, the width of the local barrier is roughly proportional to $q^{-1}$, so the rate at which the current increases with voltage is sensitive to $q$. As the voltage increases, the bands on the NT side of the barrier rise, and the local barrier due to pinning decreases, vanishing at the nominal pinned threshold $\phi_{bp}$ (or slightly before, due to incomplete pinning at finite $D_0$).

At finite temperature, there is current at any non-zero voltage because of thermionic emission. This is already included correctly in Eq. (6), within the usual approximations [1]. Calculated I-V curves at room temperature are shown in Fig. 4b. These are qualitatively similar to the usual rectifying I-V curves of planar Schottky diodes, with a sharp turn-on under forward bias and a small leakage current under reverse bias.

For comparison we show dotted curves including only thermionic emission over barriers $\phi_{bp}$ and $\phi_{b0}$, as for a pinned or unpinned planar junction. The real I-V curve falls between these two limits, but is closer to the behavior of an unpinned planar contact. Over a limited range of voltage, the I-V curve can be fitted by classic thermionic emission over an effective barrier of intermediate value 0.21 eV (middle dotted line).

Our qualitative conclusions should also apply when the metal is a wire rather than planar, although the "image charge" will have a different spatial distribution. This geometry has previously been considered for NT-NT Schottky diodes [15], but that analysis did not include any interface dipole (perhaps because $\phi_{bp} = \phi_{b0}$ in that case).

Heterojunctions have also been fabricated between NTs and semiconductors such as SiC [4]. There is a close analogy between heterojunctions and Schottky barriers [8], and we anticipate similar behavior. Band lineups at planar heterojunctions are usually controlled by alignment of the neutrality levels of the two semiconductors; but for NT-semiconductor junctions, it should be possible to shift the effective offset by treating the surface of the planar semiconductor to alter the electron affinity.

F.L. acknowledges support from the NSERC of Canada.

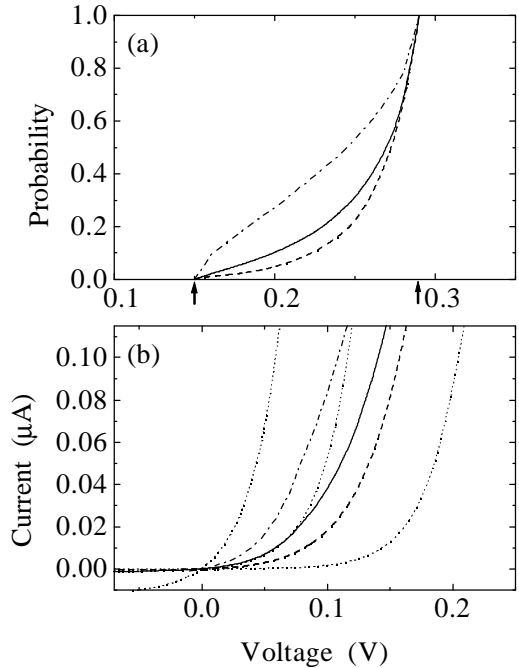

FIG. 4. (a) Tunneling probability versus voltage $V$, for NT Schottky diode as in Fig. 3b at low temperature. Arrows indicate unpinned (0.15 V) and pinned (0.29 V) barriers. Solid curve is $q = 2$ nm$^{-1}$, dashed line is $q = 1$ nm$^{-1}$, dash-dotted line is $q = 4$ nm$^{-1}$. (b) Room temperature I-V curve. Solid, dashed and dash-dotted lines are as in (a). Dotted lines are current without tunneling (as for planar contacts), for barrier heights 0.15 eV, 0.21 eV and 0.29 eV (left to right). See text for details.